\documentclass[11pt]{article}
\begin{document}

\title{
{\bf
Effects of Flavor-dependent $q\bar{q}$ Annihilation on the Mixing
Angle of the Isoscalar Octet-Singlet and Schwinger's Nonet Mass
Formula\footnote{Supported by the National Natural Science
Foundation of China under Grant
Nos. 19991487, 19677205 and 19835060, and the Foundation of the
Chinese Academy of Sciences under Grant No. LWTZ-1298.}}}

\author{ \small De-Min Li\footnote{E-mail:lidm@hptc5.ihep.ac.cn} $^a$,
~~Hong Yu$^{a,b}$~ and~ Qi-Xing Shen$^{a,b}$\\
\small $^a$Institute of High Energy Physics, Chinese Academy of Sciences,\\
\small P.O.Box $918~(4)$, Beijing $100039$, China\\
\small $^b$Institute of Theoretical Physics, Chinese Academy of Sciences, Beijing 100080, China }
\date{}
\maketitle
By incorporating the flavor-dependent quark-antiquark annihilation
amplitude into the mass-squared matrix describing the mixing of the
isoscalar states of a
meson nonet, the new version of
Schwinger's nonet mass formula which holds with a high accuracy for the
$0^{-+}$, $1^{--}$,
$2^{++}$, $2^{-+}$ and $3^{--}$ nonets is derived and the mixing angle of
isoscalar octet-singlet for these nonets is obtained. In particular, the
mixing angle of
isoscalar octet-singlet for
pseudoscalar nonet is determined to take the value of $-12.92^\circ$,
which is in agreement with the value of $-13^\circ\sim-17^\circ$ deduced from
a rather exhaustive
and up-to-date analysis of data. It is also pointed out that the
omission of
the flavor-dependent $q\bar{q}$ annihilation effect might be a factor
resulting in the invalidity of Schwinger's original nonet mass formula for
pseudoscalar nonet.\\

\vspace{1cm}

{{\bf PACS numbers:} 11.30.Hv, 12.40.Yx, 14.40-n}

\newpage
\baselineskip 24pt
\indent
According to the quark model, the $q\bar{q}$ mesons containing $u$, $d$
and $s$ quarks correspond to an octet and a singlet of the $SU(3)$ flavor
group:
\begin{equation} 3\otimes\bar{3}=8\oplus1.
\end{equation}
In general, states with the same isospin-spin-parity $IJ^{PC}$ and additive
quantum numbers can mix. Thus, the $I=0$ member of ground state octet
$\eta_8$ mixes with the corresponding singlet $\eta_1$ to produce the two
physical states $\eta$ and $\eta^\prime$. Here, we assume that the
possibility of the mixing of $\eta_8$, $\eta_1$ and other isosinglets
such as glueball and hidden flavor heavy quark meson can be ignored.

In the
$\eta_8=(u\bar{u}+d\bar{d}-2s\bar{s})/\sqrt{6}$
and $\eta_1=(u\bar{u}+d\bar{d}+s\bar{s})/\sqrt{3}$ basis, the mass-squared
matrix $M^2$ describing the mixing of $\eta_8-\eta_1$ can be
described as
\begin{equation}
M^2=\left( \begin{array}{cc}
M^2_{88}& M^2_{18}\\
M^2_{18}& M^2_{11}
\end{array}\right),
\end{equation}
where $M^2_{88}$ is the sum of the mass-squared of $\eta_8$
and the transition amplitude of $\eta_8\leftrightarrow
gg...g\leftrightarrow\eta_8$, $M^2_{11}$ is
the sum of the mass-squared of $\eta_1$ and the transition amplitude
of $\eta_1\leftrightarrow gg...g\leftrightarrow\eta_1$,
and $M_{18}$ is the transition amplitude of $\eta_8\leftrightarrow
gg...g\leftrightarrow\eta_1$ (g denotes gluon). $M^2$ satisfies
\begin{equation}
UM^2U^{-1}=\left(\begin{array}{cc}
M^2_{\eta}&0\\ 0&M^2_{\eta^\prime}
\end{array}\right),
\end{equation}
where
\begin{equation}
U=\left(\begin{array}{cc} \cos\theta&-\sin\theta\\ \sin\theta&\cos\theta
\end{array}\right),
\end{equation}
$\theta$ is the mixing angle of
$\eta_8-\eta_1$, $M_{\eta}$ and $M_{\eta^\prime}$ are the masses of
the physical states $\eta$ and $\eta^\prime$, respectively. Equation (3)
reads
\begin{equation}
M^2=U^{-1}\left(\begin{array}{cc}
M^2_{\eta}&0\\ 0&M^2_{\eta^\prime}
\end{array}\right)U=
\left(\begin{array}{cc} \cos^2\theta
M^2_{\eta}+\sin^2\theta M^2_{\eta^\prime}&
\sin\theta\cos\theta(M^2_{\eta^\prime}-M^2_{\eta})\\
\sin\theta\cos\theta(M^2_{\eta^\prime}-M^2_{\eta})&
\sin^2\theta M^2_{\eta}+\cos^2\theta M^2_{\eta^\prime}
\end{array}\right).
\end{equation}
Comparing Eq. (2) with Eq. (5), one can get the following relations:
\begin{eqnarray}
&&\tan^2\theta=\frac{M^2_{88}-M^2_{\eta}}
{M^2_{\eta^\prime}-M^2_{88}},\\
&&\sin2\theta=\frac{2M^2_{18}}
{M^2_{\eta^\prime}-M^2_{\eta}},\\
&&\tan2\theta=\frac{2M^2_{18}}
{M^2_{11}-M^2_{88}},\\
&&\tan\theta=\frac{M^2_{88}-M^2_{\eta}} {M^2_{18}}.
\end{eqnarray}
The mixing angle $\theta$ can be determined by any of the above four
relations.

In the presence of the
flavor-independent $q\bar{q}$ annihilation, the matrix elements of $M^2$
can be given by Refs.\cite{Close,Dm,Bura,Bura1}, i.e.,
\begin{eqnarray}
&&M^2_{88}=\frac{1}{3}(4M^2_K-M^2_{\pi^0}),\\
&&M^2_{18}=-\frac{2\sqrt{2}}{3}(M^2_K-M^2_{\pi^0}),\\
&&M^2_{11}=\frac{1}{3}(2M^2_K+M^2_{\pi^0}+9A),
\end{eqnarray}
where $A=(M^2_{\eta^\prime}+M^2_{\eta}-2M^2_K)/3$, $M_{\pi^0}$ and $M_K$
are the masses of the isovector $\pi^0$ and isodoublet
$K$, respectively; $M^2_K=(M^2_{K^\pm}+M^2_{K^0})/2$.
From Eqs. (2) and (5), one can get
\begin{eqnarray}
&&M^2_{88}+M^2_{11}=M^2_{\eta}+M^2_{\eta^\prime},\\
&&M^2_{88}M^2_{11}-M^2_{18}M^2_{18}=M^2_{\eta}M^2_{\eta^\prime}.
\end{eqnarray}
By eliminating $A$ from the two relations, one can get
Schwinger's original nonet mass formula\cite{Sch}
\begin{equation}
(4M^2_K-3M^2_{\eta}-M^2_{\pi^0})(3M^2_{\eta^\prime}+M^2_{\pi^0}-4M^2_K)
=8(M^2_K-M^2_{\pi^0})^2.
\end{equation}
However, Eqs.(10)$\sim$(12) are not self-consistent. First, for the $0^{-+}$
($\pi$,~$K$,~$\eta$,~$\eta^\prime$), $1^{--}$
($\rho$,~$K^\ast$,~$\omega$,~$\phi$), $2^{++}$
($a_2(1320)$,~$K^\ast_2(1430)$,~$f_2(1270)$,~$f^\prime_2(1525)$), $2^{-+}$
($\pi_2(1670)$,~$K_2(1770)$,~$\eta_2(1645)$,~$\eta_2(1870)$)  and $3^{--}$
($\rho_3(1690)$,~$K^\ast_3(1780)$,~$\omega_3(1670)$,~$\phi_3(1850)$)
nonets (All the masses of the physical states are taken from
Particle Data Group 98\cite{PDG}), the left-hand side and
right-hand
side of Eq. (15) are not
balance, especially for pseudoscalar nonet Eq. (15) is obviously invalid
(see the columns II, III and IV of Table 1). Second, for a meson nonet,
the values of the mixing angle derived from different relations
($\theta_6\sim\theta_9$) are different (see Table
2). Therefore, the matrix elements of $M^2$ ( Eqs. (10)$\sim$(12) ) should
be
modified.

Recently, Burakovsky et al.\cite{Bura, Bura1} discussed this problem
by incorporating the pseudoscalar decay constants into the matrix
elements of $M^2$, however, which is valid only for pseudoscalar nonet. In
this letter, we shall discuss the
same issue by incorporating the effect of the flavor-dependent $q\bar{q}$
annihilation into the matrix elements of $M^2$, which is valid for all
above nonets.

We assume that the transition between $q\bar{q}$ and
$q^\prime\bar{q^\prime}$ is flavor-dependent\cite{De}, i.e., the
transition between different flavor quarkonia is not flavor blind, taking
into account the possibility that the nonstrange quarkonia and
strange quarkonia system have the different wave functions at the origin
as the
result of the different mass.
In the $N=(u\bar{u}+d\bar{d})/\sqrt{2}$, $S=s\bar{s}$ basis, from
Refs.\cite{Fuch,Feld,Kawai}, the mass-squared matrix $M^2$ can be replaced by
\begin{equation}
{M^{\prime}}^2=\left(\begin{array}{cc}
M^2_N+r^2A^\prime&rA^\prime\\
rA^\prime&M^2_S+A^\prime
\end{array}\right),
\end{equation}
where $r$ describes the effect of the flavor-dependent $q\bar{q}$
annihilation, $r=\sqrt{2}$ means $q\bar{q}$
annihilation is flavor-independent; $A^\prime$ is the
transition amplitude of $S\leftrightarrow gg...g\leftrightarrow S$. Owing to
\begin{equation}
(N,S)=(\eta_8,\eta_1)R=(\eta_8,\eta_1)\left(\begin{array}{cc}
\sqrt{\frac{1}{3}}&-\sqrt{\frac{2}{3}}\\
\sqrt{\frac{2}{3}}&\sqrt{\frac{1}{3}}
\end{array}\right),
\end{equation}
the mass-squared matrixes $M^2$ and ${M^\prime}^2$ can be connected by
\begin{equation}
M^2=R{M^{\prime}}^2R^{-1}.
\end{equation}
If we assume $M_N=M_{\pi^0}$ and
$M^2_S=2M^2_K-M^2_N$\cite{Feld,Kawai,Gell},
the matrix elements of $M^2$ can now be replaced by
\begin{eqnarray}
&&M^2_{88}=\frac{1}{3}(4M^2_K-M^2_{\pi^0})+
\frac{1}{3}A^{\prime}r^2-\frac{2\sqrt{2}}{3}A^{\prime}r+\frac{2}{3}A^{\prime},
\\
&&M^2_{18}=-\frac{2\sqrt{2}}{3}(M^2_K-M^2_{\pi^0})
+\frac{\sqrt{2}}{3}A^{\prime}r^2-\frac{1}{3}A^{\prime}r-
\frac{\sqrt{2}}{3}A^{\prime},
\\
&&M^2_{11}=\frac{1}{3}(2M^2_K+M^2_{\pi^0})
+\frac{2}{3}A^{\prime}r^2+\frac{2\sqrt{2}}{3}A^{\prime}r+\frac{1}{3}A^{\prime},
\end{eqnarray}
where
\begin{equation}
A^\prime=
\frac{(M^2_{\eta^\prime}-2M^2_K+M^2_{\pi^0})(M^2_{\eta}-2M^2_K+M^2_{\pi^0})}
{2(M^2_{\pi^0}-M^2_K)},
\end{equation}
\begin{equation}
r^2=\frac{(M^2_{\eta}-M^2_{\pi^0})(M^2_{\pi^0}-M^2_{\eta^\prime})}
{(M^2_{\eta^\prime}-2M^2_K+M^2_{\pi^0})(M^2_{\eta}-2M^2_K+M^2_{\pi^0})}.
\end{equation}

Based on Eqs. (13) and (14), the new version of Schwinger's nonet mass
formula including the effect of the flavor-dependent $q\bar{q}$ annihilation
can be derived as
\begin{equation}
[2r^2M^2_K-(1+r^2)M^2_{\eta}-(r^2-1)M^2_{\pi^0}]
[(1+r^2)M^2_{\eta^\prime}+(r^2-1)M^2_{\pi^0}-2r^2M^2_K]=
4r^2(M^2_K-M^2_{\pi^0})^2.
\end{equation}
If $r=\sqrt{2}$, Eq. (24) can be reduced to Eq. (15).
From Eqs. (19)$\sim$(24), both sides of Eq. (24) are given in the
columns V and VI of Table 1, and the values of mixing angle determined
from different relations ( $\theta_6\sim\theta_9$ ) are shown in Table 3.

The columns V and VI of Table 1 show that the new version of Schwinger's
nonet mass formula holds with a high accuracy for all above nonets.
At the same time, Table 3 indicates that for a meson nonet, the values of
the mixing angle derived from different relations
($\theta_6\sim\theta_9$) are exactly equal. Furthermore,
comparing Table 2 with Table 3, one can clearly conclude that
the effect of the flavor-dependent $q\bar{q}$ annihilation on the mixing
angle of $\eta_8-\eta_1$ for
pseudoscalar nonet is quite significant while relatively weak for
the $1^{--}$, $2^{++}$, $2^{-+}$ and $3^{--}$ nonets. The
mixing angle of $\eta_8-\eta_1$ for pseudoscalar
nonet, $\theta=-12.92^\circ$, is also consistent with the value of
$-13^\circ\sim-17^\circ$ deduced
from a rather exhaustive
and up-to-date analysis of data including strong decays of tensor and
higher spin mesons, electromagnetic decays of vector and
pseudoscalar mesons, and the decays of $J/\psi$\cite{Bramon}.

It should be emphasized that the only difference between Eqs. (15)
and (24) is that Eq. (24) contains $r$, the term describing the effect
of the flavor-dependent $q\bar{q}$ annihilation. However, for pseudoscalar
nonet Eq. (24) holds with a high accuracy while Eq. (15) is obviously
invalid, which implies that the omission of the flavor-dependent
$q\bar{q}$
annihilation effect might be a factor resulting in the failure of
Schwinger's original nonet mass formula for pseudoscalar nonet.

In conclusion,
by investigating the effect of the flavor-dependent $q\bar{q}$
annihilation on the mixing angle of isoscalar octet-singlet and
Schwinger's nonet mass formula for the  $0^{-+}$, $1^{--}$,
$2^{++}$, $2^{-+}$ and $3^{--}$ nonets\footnote{The
related discussions have been done in Ref.\cite{Li} for $1^{++}$ nonet},
we find that the effect of the flavor-dependent $q\bar{q}$
annihilation should be considered  when
we discuss the mixing of isoscalar octet-singlet of a meson nonet,
especially for pseudoscalar nonet.
We believe that the omission of flavor-dependent $q\bar{q}$ annihilation
effect might be a factor
resulting in the invalidity of Schwinger's original nonet mass formula for
pseudoscalar nonet.

\newpage

\small
\begin{table}
\begin{center}
{\bf Table 1.} Values of $l_{15}$ and $r_{15}$ ( $l_{24}$ and $r_{24}$ )
denoting the left hand side and right hand side of Eq. (15) ( Eq.(24)),
respectively.

\vspace{0.3cm}

\begin{tabular}{|c|c|c|c|c|c|c|c|c|}\hline
Nonet&$l_{15}$&$r_{15}$&$|\frac{l_{15}-r_{15}}{l_{15}}|$
&$l_{24}$&$r_{24}$\\\hline $0^{-+}$&
$0.1178$&$0.4140$&$251\%$&$0.6788$&$0.6788$\\\hline $1^{--}$&
$0.3956$&$0.3399$&$14.1\%$&$0.1045$&$0.1045$\\\hline $2^{++}$&
$0.8485$&$0.7426$&$12.5\%$&$1.6049$&$1.6049$\\\hline $2^{-+}$&
$0.9455$&$1.006$&$6.4\%$&$0.8059$&$0.8059$\\\hline $3^{--}$&
$0.7815$&$0.7303$&$6.6\%$&$1.2153$&$1.2153$\\\hline \end{tabular}
\end{center}
\end{table}

\begin{table}
\begin{center}
{\bf Table 2.}
Values of the mixing angles ($\theta_6$, $\theta_7$, $\theta_8$
and $\theta_9$ are respectively derived from Eqs. (6), (7), (8) and (9) in
the presence of
the flavor-independent $q\bar{q}$ annihilation).

\vspace{0.3cm}

\begin{tabular}{|c|c|c|c|c|c|c|}\hline
Nonet&$\theta_6$&$\theta_7$&$\theta_8$&$\theta_9$\\\hline
$0^{-+}$&
$-10.87^\circ$&$-21.98^\circ$&$-18.39^\circ$&$-5.85^\circ$\\\hline
$1^{--}$&
$-50.73^\circ$&$-57.35^\circ$&$-51.16^\circ$&$-52.84^\circ$\\\hline
$2^{++}$&
$-59.34^\circ$&$-62.42^\circ$&$-60.16^\circ$&$-60.99^\circ$\\\hline
$2^{-+}$&
$-61.56^\circ$&$-60.11^\circ$&$-61.15^\circ$&$-60.80^\circ$\\\hline
$3^{--}$&
$-58.24^\circ$&$-60.04^\circ$&$-58.63^\circ$&$-59.10^\circ$\\\hline
\end{tabular}
\end{center}
\end{table}
\begin{table}

\begin{center}
{\bf Table 3.}
Values of the mixing angles ($\theta_6$, $\theta_7$,
$\theta_8$ and $\theta_9$ are respectively derived from Eqs. (6), (7),
(8) and (9) in the presence of the flavor-dependent $q\bar{q}$
annihilation).

\vspace{0.3cm}

\begin{tabular}{|c|c|c|c|c|c|c|}\hline
Nonet&$\theta_6$&$\theta_7$&$\theta_8$&$\theta_9$\\\hline
$0^{-+}$&
$-12.92^\circ$&$-12.92^\circ$&$-12.92^\circ$&$-12.92^\circ$\\\hline
$1^{--}$&
$-51.31^\circ$&$-51.31^\circ$&$-51.31^\circ$&$-51.31^\circ$\\\hline
$2^{++}$&
$-59.00^\circ$&$-59.00^\circ$&$-59.00^\circ$&$-59.00^\circ$\\\hline
$2^{-+}$&
$-61.51^\circ$&$-61.51^\circ$&$-61.51^\circ$&$-61.51^\circ$\\\hline
$3^{--}$&
$-58.12^\circ$&$-58.12^\circ$&$-58.12^\circ$&$-58.12^\circ$\\\hline
\end{tabular}
\end{center}
\end{table}

\end{document}